\documentstyle[prd,aps]{revtex} 
\begin{document} 
%
\title{Microtorsion}
\author{G.R. Filewood}
\address{School of Physics,
University of Melbourne,
Parkville, Victoria 3052 Australia.}
\maketitle
\begin{abstract}
The problem of
unification of electro-magnetism
and gravitation in four dimensions;
 some new ideas involving 
torsion. A metric
consisting of a combination of symmetric and 
anti-symmetric parts is postulated and, in
 the framework
of general covariance, used to  derive the
free-field electro-magnetic stress-energy
 tensor and the 
source tensor.
\end{abstract}
\section{Introduction}
H.
Weyl, by the use of scale transformations,
attempted to unify gravitation and electro-magnetism
within the framework of general relativity
early this century
\cite{Weyl} and in so doing initiated 
the `gauge revolution' in 
physics.
Like Einstein and many others
who followed, Weyl recognised that the 
two forces which propagate at the speed of 
light must be intimately connected.
Today we understand that
there  are profound similarities between
electromagnetism and gravitation in spite of
the obvious differences; both are gauge
field theories, both are mediated by
massless bosons (if we accept the
reality of gravitons)
and both manifest as
waves in the vacuum in classical 
(non-quantum) theory. However at present
these two
forces are described by very different
physical principles; in the case of gravitation
by a metric theory of general relativity
which relates gravity to space-time structure
whilst in the classical (and quantum) field
theory of 
electro-magnetism space-time geometry is only in
the background. Thus whilst the standard descriptions
are not contradictory they are also not 
cohesive; intuitively we feel that two such
similar forces should be based on similar
physical principles. Indeed, the general idea
that nature, at its most fundamental level
of structure, should be simple would seem
to require a single set of physical principles
to underpin these two forces lest nature be 
required to `reinvent' itself to create two 
forces based on two quite different foundations.

Thus few would dispute the need for a unified
theory but the dichotomy persists despite
nearly a century of work. The discovery of 
the weak and strong interactions has
further complicated the picture; the 
successful $SU3_{(C)}{\times}SU2_{(L)}{\times}U_1$
resisting the incorporation of the
(nonrenormalisable) Einstein Lagrangian.

Recently deAndrade and Pereira \cite{A&P}
have pointed out that, in addition to the
known result that General Relativity
can be recast into an equivalent gauge 
theory of the translation group due to teleparallel
geometry \cite{CM}\cite{YM}(leading to `dual' descriptions of 
gravitation - one describing gravitation as
propagating space-time curvature and the 
other `teleparallel' description describing
gravitation as propagating torsion), electromagnetism
additionally can have such a dual description
and that the gauge invariance of the theory
is in fact NOT violated by the couping to
torsion. This is in contradistinction to the
usual wisdom which precludes torsion coupling to 
Proca's equation for $m=0$ \cite{four} so 
that theories of torsion in electro-magnetism
usually imply photon mass. More will be said
 about this apparent conflict later - and solutions
 proposed  - but 
consider the following. If it is possible to
have `dual' descriptions of gravitation might
it be possible to have `dual' descriptions of 
electro-magnetism which in some way `complement'
gravity theory? 

Consider the  motivation for this proposition
a different way. The Coleman-Mandula (C-M or no-go
theorem) is the rock which bars the way for 
unification. This theorem forbids the (non-trivial)
union of compact groups (such as U1) and non-compact
groups (such as the Poincar\'{e} group or the
Lorentz group). However, operators
which interconvert bosons and fermions bypass the
theorem; this is the underlying motivation for
supersymmetry. This theory however requires
a whole menagerie  of superpartner particles
for which there is currently no empirical evidence. Whilst
it is thus assumed that the superpartners are 
more massive than currently accessible energies
the situation is somewhat unsatisfactory. An alternative
is highly desirable. 

 Part of 
the motivation 
for this paper is an attempt to avoid the 
C-M theorem by creating `dual' and 
complementary descriptions of  
electromagnetism in a single metric theory.
Roughly speaking what is formed is
a teleparallel version of electromagnetism
with zero non-metricity
 (curvature=0, torsion$\neq0$ for the free-field)
but with substantial differences from
previous attempts. Chief among these is the
attempt to mirror supersymmetry by
the creation of a spinorial representation
for bosonic torsion. Normally we interpret a 
metric as defining distances in space-time for 
an observer. Any component of a metric 
which defines a $ds^2=0$ component does not contribute 
to such a length; i.e. it  does not define a measurement
in an observer frame as such. 
Similarly given the equation of geodesic
or autoparallel line;
\begin{equation}
{{dx^{\alpha}}\over{ds^2}}
+
\Gamma^{\alpha}_{\beta\gamma}
{{dx^{\beta}}\over{ds}}
{{dx^{\gamma}}\over{ds}}
=0
\end{equation}
it is clear that
the torsion tensor
 $\Gamma^{\alpha}_{[\beta\gamma]}
={1\over2}
(\Gamma^{\alpha}_{\beta\gamma}
-\Gamma^{\alpha}_{\gamma\beta})
$
does not contribute since the differentials
commute. This leads to interpretation
of torsion as a non-propagating spin-contact 
interaction (see for example \cite{FP}).
It will be shown below however that a different
interpretation is possible consistent with the
findings of deAndrade and Pereira.
A  $ds^2=0$ component
is `on the light cone';
this will be used as a vacant mathematical slot into
which is plugged a spinorial
 description of electro-magnetism. 
This exploits the epicentre of the C-M
theorem; the Lorentz group (and effectively 
also the Poincar\'{e} since all distances
are zero for an `observer' on a wave of light) is non-compact
precisely because the speed of light 
is not in any observer's frame. The analogue of
a supersymmetry transformation then becomes the
interconversion of two mathematical
descriptions of one force; one spinorial and
one bosonic which is called in the text a
`translational' procedure. This `translation'
is the weakest link in the theoretical construction
but some concrete mathematical support
for its consistency is supplied by studying the construction
of stress-energy tensors from Lagrangians
with anti-symmetric metrics
in section VII. With this translation procedure
the superpartner of the photon becomes -  the
photon!; but this superpartner is only detectable
to an observer who is travelling at the speed of
light so it never appears in experiments!

What are the problems? Different people 
might give different answers to this
question but the following are a selection of
the main obstacles to four-dimensional
geometric unification of electro-magnetism
and gravitation in the framework of
general relativity;

1. For a dimensionless metric the 
scalar curvature R has dimension $l^{-2}$
so that in four dimensions ${\cal{L}}=kR$
requires constant k to have dimension
$l^{-2}$ and so the theory is non-renormalisable.

2. How can the electro-magnetic vector
potential $A_{\mu}$ be placed in the
tensor $g_{\mu\nu}$ without spoiling
its tensor character or destroying
General Relativity; i.e. if  we require the
strong equivalence principle
$g_{\mu\nu\;;\;\phi}=0$ to hold. Note
that this is related to the first problem
because $A_{\mu}$ is dimensional with
dimension $l^{-1}$.

3. How can the free-field stress-energy tensor
be extracted from the metric? From the
Lagrangian?

4. How can source terms be included in
the metric? Again these must not
spoil the qualities of the metric
or destroy gravitation theory.

5. How can we couple the stress-energy
tensors for gravitation and electro-magnetism
into one equation relating to curvature?

6. Does the theory have scope for 
generalisation to the electro-weak
interaction?

7. And what about quantisation?

8. The Coleman-Mandula  theorem.

The paper is organised as follows;
firstly there is a brief review of
the history - particularly
regarding homothetic
curvature (\'{a}-l\'{a}-Eddington)
with which many readers may not be
familiar. A metric is then defined and its
consistency proven. It contains
both a symmetric and an anti-symmetric
part. A connection is then
defined on the basis of the vanishing of the
covariant derivative of the metric
(vanishing non-metricity). 
The stress-energy tensor is then extracted 
by expanding the (homothetic) curvature tensor in the
form of an Einstein equation. 
The metric is then redefined to accomodate
source terms and the source-stress-energy
tensor formed. Lastly the  metric
is applied to the Lagrangian formalism.
Due to constraints of space
the  prospects for electro-weak-gravitation
 unification and explicit interaction terms
are not
discussed. The notation used throughout
is perhaps somewhat traditional as
the particular mathematical structure explored
does not lend itself ideally to the
notation of differential forms (e.g.
 the use of notation with $\omega$ connection one-forms,
exterior derivative, exterior product etc; see
for example Trautman \cite{T}) because every
index must be carefully tracked for anti-commutivity.
The notation used is consistently applied
and certainly familiar to anyone accustomed
to the standard texts on General Relativity.
Part of the work is an extension of ideas
presented previously
\cite{GF}.

There is an extensive literature in this
field but the ideas proposed in this paper
are quite different from any previously published
work to the best of my knowledge.

\section{Historical background}
It is instructive to 
consider the original efforts since
the principles uncovered by the pioneers
in the field underpin all efforts that
followed.
Theoretical efforts to form a unified
description of gravity and electro-magnetism
in the classical  framework date from
early this century beginning particularly
with the work of  N\"{o}rdstrom, Weyl, Eddington,
 Einstein and
Cartan. An excellent review is found
in reference \cite{russian}.
 Weyl's scheme \cite{Weyl} revolved around
scale-transformations (gauge transformations)
. This work failed to provide a viable
unified theory of gravitation and
 electro-magnetism, which was Weyl's original intent,
but subsequently proved very fruitful in other 
areas; Weyl is truely the father of the
modern approach of gauge field theory.
 
In essence Weyl's idea was to extend the 
geometric foundations of Riemannian 
geometry by allowing for scale transformations 
to vectors with parallel transport. This 
approach was criticised, particularly by
Einstein, as being incompatible with
observation; in particular it means that
that the physical properties of
measuring rods and clocks 
depends upon their history. For example, 
two identical clocks, initially synchronised
to run at the same rate in the same inertial
frame, will no longer do so if they are brought
together at a later time into the same 
inertial frame having travelled through
different paths in space-time according to
Weyl's scheme.
  
Subsequently
Eddington \cite{Eddington} attempted to extend
Weyl's theory. In Eddington's theory,
as in Weyl's,
the connection forms the basic geometric
object and is related to the electromagnetic
potential $A_\beta$;

\begin{equation}
\Gamma^{\alpha}_{\alpha\beta}
=
\Gamma^{\alpha}_{\beta\alpha}
=\lambda.A_{\beta}
\end{equation}
-for constant dimensionless lambda. 
Eddington then proceeds to form the anti-symmetric
electro-magnetic tensor $F_{\beta\gamma}$ 
 viz homothetic curvature;

\begin{eqnarray}
R^\alpha_{\alpha\beta\gamma}
&=&
\partial
_{\beta}\,\Gamma^{\alpha}_{\alpha\gamma}
-\partial_{\gamma}\,\Gamma^\alpha_{
\alpha\beta}
-\Gamma^{\alpha}_{\phi\beta}\,\Gamma^
{\phi}_{\alpha\gamma}+
\Gamma^{\alpha}_{\phi\gamma}\,
\Gamma^{\phi}_{\alpha\beta}
\nonumber\\
&=&
F_{\beta\gamma}
\label{dotty}
\end{eqnarray}
-where the two product terms in (\ref{dotty})
have been equated to zero as is usually done in general
relativity. In order to overcome the
criticism of Weyl's theory with regard to 
measuring rods and clocks Eddington imposes
an assumed  metric condition on the curvature tensor
(Eddington's `natural gauge');
\[
{\phi}g^{\alpha\beta}
=
R^{\alpha\beta}
\]
where $\phi$ is a constant
of dimension $l^{-2}$. This constraint
effectively `fine-tunes' the metric to the
space-time curvature in an attempt to
avoid the problem of measurement 
associated with the Weyl 
non-metric geometry.

 There are a number of problems associated
with the Eddington approach which
emerged. In particular these involve the number of
unknowns in the differential equations
resulting from the use of
the connection as the main geometric
element (about 40) and higher-order
derivative terms which arise in the 
theory.
More generally, we can see an inconsistency with
general relativity because the curvature 
tensor is equated viz Einstein's 
equation to the stress-energy tensor in G.R;
the corresponding object in electro-magnetism
is quadratic in the $F_{\mu\nu}$ not first-order
 in it. In fact it is
the E-M stress-energy tensor which
should appear on the R.H.S. of
Einstein's equation contributing, at the
very least, to the gravitational
potential as it is a source of
mass-energy equivalence. Also in the Weyl/Eddington theory
the potential is identified with the connection;
in G.R. it is identified with the metric. 
More recent studies using the Weyl/Eddington
approach are found in refs \cite{WE}.

  Cartan appears to have been the first
person to explore the possibility of theories
involving torsion in the context of general
covariance and classified possible theories
 on the basis of
affine vs metric, 
(affine theories, such as Weyl's, are 
`non-metric'), the presence or
absence of rotation curvature
(defined as present if
$R^{\alpha}_{\gamma\beta\alpha}
=R^{\alpha}_{\beta\gamma\alpha}\;
{\neq}\;0$), the presence or
absence of homothetic curvature
(
present if $R^{\alpha}_{\alpha\beta\gamma}
=-R^{\alpha}_{\alpha\gamma\beta}\;
{\neq}\;0$) and the presence or
absence of torsion
(presnt if $\Gamma^{\alpha}_{\beta\gamma}
=-\Gamma^{\alpha}_{\gamma\beta}\;
{\neq}\;0$). 
Riemann-Cartan geometries 
have non-vanishing torsion. 
Many R-C geometries involve adding an
anti-symmetric piece to the
metric which is in some way related
to the Maxwell tensor $F_{\alpha\beta}$;
\[
g_{\alpha\beta}= \eta_{\alpha\beta}
+h_{\alpha\beta}+{\lambda}F_{\alpha\beta}
\]
where $h_{\alpha\beta}$ is the gravitational
potential in the weak field 
approximation. It is now understood
that torsion is related to translations
\cite{Hehl}
(torsion `breaks' parallelograms -
it is also related to the theory of 
crystal dislocations \cite{Bilby})
whilst rotation curvature is related to
rotations. This situation is somewhat
paradoxical (as has been noted) since
the (non-compact) Poincar\'{e} group,
the `gauge' group for gravitation, 
is the group of translations.
In the treatment given here we will see 
the reverse  by embedding 
electro-magnetism (a U1 or compact 
rotation group symmetry) in torsion;
which has the geometry
of translations not rotations!
An attempt to give a geometric interpretation
to these apparent contradictions will be given
in the discussion section when all the
geometric machinery needed has been developed.

At about the same time as Eddington
published his theory Kaluza published his 
five-dimensional version of 
gravitational-electromagnetic unification
\cite{KK}.

The fifth dimension in the Kaluza-Klein theory
is a periodic space which spontaneously
compactifies. (More contemporary versions of 
the Kaluza-Klein geometry attempt to extend
the compactified space to higher dimensions
to accomodate the $S.U._3\mbox{x}S.U._2\mbox{x}
U_1$ standard model; see ref. \cite{KK} for
examples. See also \cite{F.W.}). The Kaluza-Klein theory
is appealing for a number of basic reasons.
Often overlooked but of basic importance is the fact
that the metric in the theory parallels general
relativity by containing the potential of the 
theory; most alternative attempts at unification
have attempted to site the vector potential $A_\mu$
in the connection.
However, the Kaluza-Klein theory remains a five-dimensional
theory and ideally we would like a four-dimensional
theory;
 the universe
is  not observed to be anything other than 
four dimensional so we have no empirical 
evidence for the extra dimensions. 
In addition to the Kaluza-Klein theory there are
numerous theories 
identifying an anti-symmetric component of the
metric with the Maxwell tensor $F^{\mu\nu}$
such as Einstein's unified field theory
\cite{UFT} and later contributions to $U_4$
theory (with torsion) development from
Sciama \cite{3} and Kibble \cite{4} 
and others. More 
contemproary approaches
include 3D Riemann-Cartan geometry
with Yang-Mills fields (\cite{Mielke}
\cite{Rad} and contained references).
Lunev \cite{Lunev} develops an
Einstein equation for Yang-Mills
fields but the approach used differs
from the one employed in this paper
by placing the potential in the
connection.
Garcia deAndrade and Hammond \cite{Ham}
employ the Eddington approach of equating
homothetic curvature and the Maxwell 
tensor and interpret massive torsion
quanta as massive photons.
More recently Unzicker\cite{U} has 
explored teleparallel geometry and
electromagnetism although the latter 
approch is very different from the one presented
here.
 
Attempts to embed a description of 
electro-magnetism in general covariant
theory have difficulty because, unlike gravity
which can be `transformed away' {\it locally}
in a free-fall frame, the electro-magnetic
field cannot be `transformed away' by a Lorentz 
transformation. There appears however to be 
a loophole that can be exploited here without
explicitly breaking Lorentz invariance; that of 
working on the light-cone itself - in a non-observer
frame. What this means will be discussed below. 

 It has been pointed out
\cite{four} that 
gauge freedom in Proca's equation for $m=0$
precludes torsion in electro-magnetism (but not
in the case $m\neq0$ for a spin-1 field). Thus
theories with torsion in electro-magnetism
frequently imply photon mass \cite{FP}. However,
in the derivation presented it will shown that
a constraint emerges from the connection which
permits the description of torsion in 
electro-magnetism with massless photons;
possibly providing an explanation for the
apparent conflict between the results of Hehl et. al.
and deAndrade et. al. alluded to above.

\section{Metric}

Consider the following metric;
\begin{equation}
\left({\begin{array}{cccc}
+I_4&{i\over2}\sigma_{01}&{i\over2}\sigma_{02}&
{i\over2}\sigma_{03}\\{i\over2}\sigma_{10}&-I_4&
{i\over2}\sigma_{12}&{i\over2}\sigma_{13}\\
{i\over2}\sigma_{20}&{i\over2}\sigma_{21}&-I_4&
{i\over2}\sigma_{23}\\{i\over2}\sigma_{30}&
{i\over2}\sigma_{31}&{i\over2}\sigma_{32}&-I_4\
\end{array}}\right)
=
I_4.\eta_{\alpha\beta}+
{i\over2}\sigma_{\alpha\beta}
\label{sigmametric}
\end{equation}
\\
where $
\sigma^{\alpha\beta}=
{i\over2}\left[\gamma^{\alpha},\gamma^{\beta}
\right]$
and introducing the notation $ 
\sigma^{'\alpha\beta}
={1\over2}\sigma^{\alpha\beta}
$ we have (noting in the sum 
$\sigma_{\alpha\phi}^{'}\sigma^{'\phi}
_{\;\;\;\beta},\;\phi$ can only take two
values as $\alpha$ and $\beta$ are different valued);
\begin{eqnarray}
g_{\alpha\phi}\bar{g}^\phi_{\,\,\,\beta}
&=&\left(I_4.\eta_{\alpha\phi}
+i\sigma^{'}_{\alpha\phi}\right)\left(
I_4.\eta^\phi_{\,\,\,\beta}-i\sigma^{'\phi}
_{\,\,\,\,\beta}\right)\nonumber\\&=&
\left(I_4.\eta_{\alpha\beta}
+i\sigma^{'}_{\alpha\beta}\right)
=g_{\alpha\beta}
\label{one}
\end{eqnarray}
where $\bar{g}$ is the complex-conjugate
transpose ($\dagger$)
or `dual' metric
viz; 
\begin{equation}
\left(\sigma^{'}_{\alpha\beta}
\right)^\dagger=
{-i\over4}\left[\gamma_\alpha,
\gamma_\beta\right]^\dagger=
\sigma^{'\alpha\beta}
\label{phase}
\end{equation}
so that 
\begin{equation}
\left
(g_{\alpha\beta}\right)^\dagger
=\bar{g}^{\alpha\beta}=\left(I_4.\eta^
{\alpha\beta}-i\sigma^{'\alpha
\beta}\right)
=g^{\beta\alpha}
\label{inverse}
\end{equation}
Note that;
\begin{equation}
g_{\alpha\phi}\;g_{\beta}^{\;\;\;\phi}
=g_{\alpha\beta}
\label{consistencyconstraint}
\end{equation}
is exceedingly constraining. The general solution
for the Lorentz tangent space metric (+,-,-,-)
includes an anti-symmetric component.
Now the Dirac gamma matrices do not transform as
four-vectors. However, the sigma matrices formed
from them transform as tensors;
\begin{equation}
{i\over2}\sigma_{\alpha\beta}
=
{-1\over4}[\gamma_{\alpha},\gamma_{\beta}]_{-}
\label{commutator}
\end{equation}
which transforms as a true tensor. It is the antisymmetric
version of the fundamental tensor which can
be defined as;
\begin{equation}
\eta_{\alpha\beta}=
{1\over2}\{\gamma_{\alpha},\gamma_{\beta}\}_{+}
\label{anticommuator}
\end{equation}
Which dictates the solution space for 
(\ref{consistencyconstraint})
for which metric (\ref{sigmametric}) and its
dual are the general 
solution.
Of course the anti-symmetric piece does not span
the space S.0.(3,1) at all but is confined to
the end point of boosts - i.e. $ds^2=0$  - {\it{which
is not in the group}}. The S.0.(3,1) group is
non-compact precisely because the velocity of light,
the end-point of boosts, is not in the group. What metric
(\ref{sigmametric}) does is insert a new part of the metric
onto the light cone. The transformations associated
with using the sigma matrix as the fundamental
tensor constitute a new group confined to be
on the light cone itself. Let us call this group C.0.(3,1).
We will study its properties later but we note that;
\[
\mbox{S.O.}(3,1)\;\cup\;{\mbox{C.O.}}(3,1)
\]
will be (trivially) topologically  compact.

Of all possible anti-symmetric pieces which may be
added to the metric $\eta_{\alpha\beta}$ 
only one possible term satisfies both
the consistency equation (eq.(\ref{consistencyconstraint}))
and Lorentz covariance
and it is, not suprisingly, the fundamental tensor formed  
by substituting the commutator for the anticommutator
of the gamma matrices. There is a profound underlying
symmetry in this which I do not fully understand. It
should be apparent, however, that the metric 
(\ref{sigmametric}) is
highly non-trivial and unique.

{\bf From now on I will
drop the prime on the sigma matrices
 it being understood that {\it all}
subsequent expressions
containing sigma matrices are of 
the primed form (i.e. the
factor of $1\over2$ is absorbed
into the definition).} The off-diagonal elements
of this metric are 4x4 matrices so
 the metric is 16x16. Diagonal
identities are 4x4 and each space-time
index can be  multiplied into
a 4x4 identity to couple to the
metric. Thus although the
metric is now a 16x16 matrix
we still only need four
parameters to describe the space
which is, physically and mathematically,
 still 
therefore four dimensional; as
implied by the R.H.S. of (\ref{sigmametric})
where the Greek indices range over 4 values.
Normally we expect the antisymmetric
$\sigma$ matrices to contract  
against spinors - but here they
will be contracted against
commuting co-ordinates. 
Apart from a brief comment
at the end of the paper I will not deal
further with the issue of co-ordinates.
The theory is constructed in a co-ordinate 
independent fashion.
Contraction with the dual  metric
produces the scalar identity (omitting
the implicit matrix multiplier of $I_{4}$);
\[
g_{\alpha\beta}\,\bar{g}^{\beta\alpha}
=+4\,-3=+1 
\]
although the metric is not invertible as 
a Kronecker delta.
Consequently care is required
in raising and lowering indices (see below).
The apparent lack of invertibilty of the
metric causes no problem; the 
different parts of the metric label different
fields and indices for each field are
appropriately raised and lowered with
with each field's respective metric with
cross terms generating interactions. Thus when
the physical content of the theory is inserted
the metric is well behaved.
Note that Lorentz scalars
such as $P^2=m^2$ and $ds^2$
are still invariant under this metric. This is 
important because we wish to construct a theory
which preserves physical measurements (one of the
main criticisms of the Weyl approach was that
it did not preserve physical measurement
invariants in different regions of space
\cite{russian}).
 To obtain a
dynamical theory we will require the derivatives
of the off-diagonal anti-symmetric
part of metric (\ref{sigmametric})
to be non-vanishing. To facilitate this
we introduce a parameter $|P|$,
with non-vanishing space-time derivatives
and modulus unity,
and incorporate $|P|$ into the 
sigma matrices;
\begin{equation}
\sigma_{\alpha\beta}\equiv|P|
\sigma_{\alpha\beta}
\label{|P|}
\end{equation}
We will take $|P|$ as a one-parameter group
$|P|=e^{{\pm}ik{\cdot}x}$ 
where $k^{\mu}$ is the photon four-momentum and 
$x_{\mu}$ the
space-time four-vector in units $\hbar=c=1$.
We will see below that consistency of the
metric can be maintained with this
added phase-factor.

\section{Connection coefficients}
   
We will require the vanishing of
the covariant 
derivative of the metric; 
\begin{equation}
g_{\alpha\beta;\gamma}=\left
(I_4.\eta_{\alpha\beta}+
i|P|\,\sigma_{\alpha\beta}\right)_
{;\gamma}=0.
\label{metric}
\end{equation}
We consider only a free-fall frame 
in which the 
derivatives of the diagonal elements
of the metric
vanish; the derivatives of off-diagonal 
elements
however will
be non-vanishing in this
frame (as we shall see this applies 
when there is an
electro-magnetic field present). Now
\begin{equation}
\left(
|P|\sigma_{\alpha\phi}
\right)_{,\gamma}\,|P|\,\sigma^
{\phi}_{\;\beta}
\approx
i{|P|}_{,\gamma}
\sigma_{\alpha\beta}
=i\left({|P|}
\sigma_{\alpha\beta}
\right)_{,\,\gamma}
\label{P}
\end{equation}
($\approx$ here
means equal
up to a (local) phase factor).
From now on the parameter 
$|P|$ will be absorbed
into the definition of the
sigma matrices
($|P|\sigma_{\alpha\beta}\equiv
\sigma_{\alpha\beta}$) in all expressions.

For both indices downstairs I will
use $e^{+ik{\cdot}x}$ and for the dual
with both indices upstairs the
$e^{-ik{\cdot}x}$ so that 
\[
(g_{\alpha\beta})^{\dagger}=\bar{g}^{\alpha\beta}
\]
Raising or lowering a single index thus eliminates the
phase factor as a consequence.

 Writing and defining the covariant 
derivative of the asymmetric 
part of the metric with 
three different labellings;
\begin{equation}
g^A_{\alpha\beta;\gamma}=g^
A_{\alpha\beta,\gamma}-
\Gamma_{\gamma\alpha}^\phi
\,i\sigma_{\phi\beta}-
\Gamma_{\gamma\beta}^\phi\,i
\sigma_{\alpha\phi}
\label{alpha}
\end{equation}
\begin{equation}
g^A_{\gamma\beta;\alpha}=g^
A_{\gamma\beta,\alpha}-
\Gamma_{\alpha\gamma}^\phi
\,i\sigma_{\phi\beta}-
\Gamma_{\alpha\beta}^\phi
\,i\sigma_{\gamma\phi}
\label{beta}
\end{equation}
\begin{equation}
g^A_{\alpha\gamma;\beta}=g^
A_{\alpha\gamma,\beta}-
\Gamma_{\beta\alpha}^\phi\,i
\sigma_{\phi\gamma}-
\Gamma_{\beta\gamma}^\phi\,i
\sigma_{\alpha\phi}
\label{gamma}
\end{equation}  
\begin{equation}
\Gamma_{\alpha\beta}^\phi={1\over2}
\left(g_{\alpha\epsilon,\beta}+g_
{\epsilon\beta,\alpha}
-g_{\alpha\beta,\epsilon}\right)
\bar{g}^{\epsilon\phi}
\label{zot}
\end{equation}
using (\ref{zot}) and raising indices
with the
anti-symmetric part
of (\ref{inverse}) (we have a choice in
this situation of raising indices either
with the symmetric part of the metric 
{\it or}, the antisymmetric part or both. For the
free-field (no interactions) we require
only the anti-symmetric part of the
metric which means, because the derivatives of the
diagonal part vanish, we are effectively
working with a purely anti-symmetric metric in the 
derivation), we have
(\ref{gamma}) + (\ref{beta})
- (\ref{alpha}) gives;
\begin{eqnarray}
g_{\alpha\gamma\,,\,\beta}
&+&g_{\gamma\beta\,,\,\alpha}
-g_{\alpha\beta\,,\,\gamma}
\nonumber\\
&=&i\sigma_{\alpha\gamma\,,\,\beta}
+i\sigma_{\gamma\beta\,,\,\alpha}
-i\sigma_{\alpha\beta\,,\,\gamma}
\label{proof}
\end{eqnarray}
provided we define
contractions on the derivative index
{\it {from its right}} as;
\begin{equation}
\sigma_{\alpha\beta\,,}{}^{\phi}
\sigma_{\phi\gamma}=
+i\sigma_{\alpha\beta\,,\,\gamma}
\label{deriv}
\end{equation}
showing (\ref{zot}) is consistent.
Notice that the connection so defined
is  antisymmetric in its lower
two indices; this 
is a torsion connection. Also note that
although $\Gamma^{\phi}_{\alpha\,\beta}
=-\Gamma^{\phi}_{\beta\,\alpha}$
we cannot use this to interchange indices
and sum connection components; $\sigma_{\alpha
\,\beta\,,\,\epsilon}\neq-\sigma_{\epsilon\,\beta
\,,\,\alpha}$ for individual components.

\section{Homothetic curvature;
$R^{\alpha}_{\alpha\beta\gamma}$}

In contrast to gravitational theory 
the homothetic  curvature is non-zero. We
contract
over the first upper and first lower
index of the curvature
tensor \cite{two}\cite{Eddington}
\begin{equation}
R^\alpha_{\alpha\beta\gamma}=\partial
_{\beta}\,\Gamma^{\alpha}_{\alpha\gamma}
-\partial_{\gamma}\,\Gamma^\alpha_{
\alpha\beta}
-\Gamma^{\alpha}_{\phi\beta}\,\Gamma^
{\phi}_{\alpha\gamma}+
\Gamma^{\alpha}_{\phi\gamma}\,
\Gamma^{\phi}_{\alpha\beta}
\label{dot}
\end{equation}
which is anti-symmetric in its two
uncontracted indices. For a 
theory of electro-magnetism we require 
first derivatives of the 
potential terms. Thus we are interested
in the product of connection 
coefficients in (\ref{dot}) which, for
the case at hand,
are non-vanishing in the 
presence of metric (\ref{metric}).
We will later see that 
the other two terms with second derivatives
of the metric cancel in (\ref{dot}).
Now consider the Bianci identity;
\begin{equation}
R^\alpha_{\alpha\beta\gamma\,;\,\delta}
+R^\alpha_{\alpha\delta\beta\,;\,\gamma}+
R^\alpha_{\alpha\gamma\delta\,;\,\beta}=0
\label{bianci}
\end{equation}
and contract with the full metric;
\begin{equation}
\left(R^\alpha_{
\alpha\beta\gamma\,;\,\delta}
+R^\alpha_{\alpha\delta\beta\,;\,\gamma}+
R^\alpha_{\alpha\gamma\delta\,;\,\beta}\right)
\bar{g}^{\beta\gamma}
=0
\end{equation}
relabelling and using the fact that
 a product of symmetric and 
anti-symmetric parts with the same 
indices is zero we obtain;
\begin{eqnarray}
-i
\left(
R^\alpha_{\alpha\beta\gamma\,;\,\delta}
-2R^\alpha_{\alpha\beta\delta\,;\,\gamma}
\right)
\sigma^{\beta\gamma}&=&0\nonumber\\
{1\over2}R^{A}_{\,\,\,;\,\delta}-R^\gamma_{\,\,\,
\delta\,;\,\gamma}
&=&0
\label{einstein's}
\end{eqnarray}
where the scalar $-i
\,R^\alpha_
{\alpha\beta\gamma;\,\delta}
\sigma^{\beta\gamma}
\equiv R^{A}_{\,\,\,\,;\,\delta}$
and indices are contracted in the tensor part. 
Finally relabelling and raising indices with
the {\it symmetric} part of the metric
we obtain;
\begin{equation}
\left({1\over2}\eta^{\phi\delta}R_A-R^{\delta\phi}
\right)_{;\,\delta}=0
\label{delta}
\end{equation}
Although this equation appears identical to Einstein's
equation it contains very different information.
 Note also that
in (\ref{delta}) I have employed the opposite sign
convention than is usual in Einstein's equation. This
is a reasonable assertion since
the gravitational potential is unbounded
from below whilst the electro-magnetic potential
for a charged object is unbounded from above
as $r\rightarrow0$ so we expect curvatures which enter
with opposite sign.
%

\section{calculation of electro-magnetic torsion}

Using (\ref{inverse}), (\ref{zot})
and (\ref{deriv}) we have;
\begin{equation}
\Gamma^{\alpha}_{\alpha\beta}
={1\over2}g_{\alpha\epsilon,\beta}\,\bar{g}^
{\epsilon\alpha}
=-\Gamma^\alpha_{\beta\alpha}
\end{equation}
and thus;
\begin{eqnarray}
&\partial&
_{\beta}\,\Gamma^{\alpha}_{\alpha\gamma}
-\partial_{\gamma}\,\Gamma^\alpha_{
\alpha\beta}
\nonumber\\
&=&
{1\over2}g_{\alpha\epsilon,\gamma,\beta}
\bar{g}^{\epsilon\alpha}
+{1\over2}g_{\alpha\epsilon,\gamma}
\bar{g}^{\epsilon\alpha}_{\,\,\,,\beta}
-{1\over2}g_{\alpha\epsilon,\beta,\gamma}
\bar{g}^{\epsilon\alpha}
-{1\over2}g_{\alpha\epsilon,\beta}
\bar{g}^{\epsilon\alpha}_{\,\,\,,\gamma}
\nonumber\\
&=&
{1\over2}g_{\alpha\epsilon,\gamma}
\bar{g}^{\epsilon\alpha}_{\,\,\,,\beta}
-{1\over2}\bar{g}_{\epsilon\alpha,\beta}
g^{\alpha\epsilon}_{\,\,\,,\gamma}
\nonumber\\
&=&0
\end{eqnarray}
where the last line follows because the
$g_{\epsilon\alpha}$'s
commute as do the derivative indices.
Hence the components containing derivatives of 
the connection
of ({\ref{dot}) vanish and  we have;
\begin{eqnarray}
&-&i
R^\alpha_{\alpha\gamma\beta}
\sigma^{\gamma\beta}=
-i\left(
\Gamma^\alpha_{\phi\beta}\,\Gamma^\phi_{\alpha\gamma}-
\Gamma^\alpha_{\phi\gamma}\,\Gamma^\phi_{\alpha\beta}
\right)\sigma^{\gamma\beta}
\nonumber\\
&=&
-2i
\Gamma^\alpha_{\phi\beta}\,\Gamma^\phi_{\alpha\gamma}
\sigma^{\gamma\beta}
\nonumber\\
&=&
-{i\over2}
\left(
\begin{array}{c}
i\sigma_{\phi}{}^{\alpha}{}_{,\,\beta}\\
\mbox{\tiny(A)}
\end{array}
\begin{array}{c}
+i\sigma^{\,\alpha}_{\;\;\;\beta\,,\,\phi}\\
\mbox{\tiny(B)}
\end{array}
\begin{array}{c}
-i\sigma_{\phi\,\beta\,,}^{\;\;\;\;\;\;\alpha}\\
\mbox{\tiny(C)}
\end{array}
\right).
\nonumber\\
&\;&\;
\left(
\begin{array}{c}
+i\sigma_{\alpha}{}^{\phi}{}_{,\,\gamma}\\
\mbox{\tiny(D)}
\end{array}
\begin{array}{c}
+i\sigma^{\phi}{}_{\gamma\,,\,\alpha}\\
\mbox{\tiny(E)}
\end{array}
\begin{array}{c}
-i\sigma_{\alpha\,\gamma\,,}^{\;\;\;\;\;\;\phi}\\
\mbox{\tiny(F)}
\end{array}
\right)\sigma^{\gamma\beta}
\label{array}
\end{eqnarray}
Now consider the product involving terms (B) and (E);
\begin{eqnarray}
-{i\over2}
i\sigma^{\,\alpha}_{\;\;\;\beta\,,\,\phi}\:
i\sigma^{\,\phi}_{\;\;\;\gamma\,,\,\alpha}
\sigma^{\gamma\beta}
&=&
-{1\over2}\sigma^{\,\alpha\beta\,,\,\phi}\:
\sigma_{\phi\beta\,,\,\alpha}
\nonumber\\
&
\stackrel
{\mbox{(def.)}}
{\equiv}&
-{1\over2}\partial^{\phi}A^{\alpha}\partial_{\alpha}
A_{\phi}
\label{Adefinition}
\end{eqnarray}
The last line involves a contraction over
$\beta$ and a 
dimensional transmutation 
to define the A field. This
definition is the
`translation' alluded to earlier in the
paper and is discussed extensively
later in the paper.
Similarly
the product of terms (C) and (F)
of eq (\ref{array}) gives an identical
$-{1\over2}\partial^{\phi}A^{\alpha}\partial_{\alpha}
A_{\phi}$. For (B).(F) and (C).(E)
of eq. (\ref{array}) we
obtain;
\[
-{i\over2}
\sigma^{\,\alpha}_{\;\;\;\beta\,,\,\phi}
\sigma_{\alpha\,\gamma\,,}^{\;\;\;\;\;\;\phi}
\sigma^{\gamma\beta}
\equiv
+{1\over2}\partial^{\phi}A^{\alpha}\partial_{\phi}
A_{\alpha}
\]
The products (B).(D), (C).(D)
are zero because the
$\sigma^{\gamma\beta}$
commutes past the derivative index of
$\sigma^{\phi}{}_{\alpha\,,\,\gamma}$
and hence contracts with opposite
sign on the $\gamma$ and
$\beta$. The products (A)(E) and
(A)(F) are also zero for the same
reason (to see this first anti-commute
the two matrices;
$
\sigma_{\phi\,,\,\beta}^{\;\;\alpha}\,
\sigma^{\,\phi}_{\;\;\;\gamma\,,\,\alpha}
$
- note also that two sigma's
with dummy contracted indices anti-commute 
if, with relabelling, there is only one
index interchange on the sigma's - otherwise 
they commute).
The last product term ((A).(D) in eqn. (\ref{array}))
is zero because the derivative indices commute. 
Hence we have for the scalar
part of (\ref{delta}) summing contributions
R equals;
\begin{equation}
-\partial^{\phi}A^{\alpha}\partial_{\alpha}
A_{\phi}
+\partial^{\phi}A^{\alpha}\partial_{\phi}
A_{\alpha}
=
{1\over2}F^{\phi\alpha}F_{\phi\alpha}
\end{equation}
The tensor part of (\ref{delta}) is similarly
calculated. A subtlety however arises with regard to
translations into forms like (\ref{Adefinition}) because
of the anti-symmetry of the tensor piece $R_{\gamma}^{\;\;\delta}$.
I will calculate the terms first, impose anti-symmetry on the 
translation into the A-field terms `by hand' and then explain the
meaning of the translation later in the text;
\begin{eqnarray}
&+&2i
R^\alpha_{\alpha\delta\beta}\,
\sigma^{\gamma\beta}
=
+2i\left(
\Gamma^{\alpha}_{\phi\beta}\,
\Gamma^{\phi}_{\alpha\delta}-
\Gamma^{\alpha}_{\phi\delta}\,
\Gamma^{\phi}_{\alpha\beta}
\right)\sigma^{\gamma\beta}
\nonumber\\
&=&+
{i\over2}
\left(
\begin{array}{c}
\mbox{\tiny(A)}\\
i\sigma_{\phi}{}^{\alpha}{}_{,\,\beta}
\end{array}
\begin{array}{c}
\mbox{\tiny(B)}\\
+i\sigma^{\,\alpha}_{\;\;\;\beta\,,\,\phi}
\end{array}
\begin{array}{c}
\mbox{\tiny(C)}\\
-i\sigma_{\phi\,\beta\,,}^{\;\;\;\;\;\;\alpha}
\end{array}
\right).
\nonumber\\
&\;&\;
\left(
\begin{array}{c}
\mbox{\tiny(D')}\\
i\sigma_{\alpha\;\;\,,\,\delta}^{\;\;\phi}
\end{array}
\begin{array}{c}
\mbox{\tiny(E')}\\
+i\sigma^{\,\phi}_{\;\;\;\delta\,,\,\alpha}
\end{array}
\begin{array}{c}
\mbox{\tiny(F')}\\
-i\sigma_{\alpha\,\delta\,,}^{\;\;\;\;\;\;\phi}
\end{array}
\right)\sigma^{\gamma\beta}
\nonumber\\
&\;&\;-
{i\over2}
\left(
\begin{array}{c}
i\sigma_{\phi\;\;\,,\,\delta}^{\;\;\alpha}\\
\mbox{\tiny(A')}
\end{array}
\begin{array}{c}
+i\sigma^{\,\alpha}_{\;\;\;\delta\,,\,\phi}\\
\mbox{\tiny(B')}
\end{array}
\begin{array}{c}
-i\sigma_{\phi\,\delta\,,}^{\;\;\;\;\;\;\alpha}\\
\mbox{\tiny(C')}
\end{array}
\right).
\nonumber\\
&\;&\;
\left(
\begin{array}{c}
i\sigma_{\alpha}{}^{\phi}{}_{,\,\beta}
\\
\mbox{\tiny(G)}
\end{array}
\begin{array}{c}
+i\sigma^{\,\phi}_{\;\;\;\beta\,,\,\alpha}\\
\mbox{\tiny(H)}
\end{array}
\begin{array}{c}
-i\sigma_{\alpha\,\beta\,,}^{\;\;\;\;\;\;\phi}\\
\mbox{\tiny(I)}
\end{array}
\right)\sigma^{\gamma\beta}
\label{array2}
\end{eqnarray}
The only products which are zero in
(\ref{array2}) are (A).(D') and (A').(G).
Relabelling dummies shows that the 
remaining products in 
(\ref{array2}) anti-commute.
For (B').(H) we have; 
\begin{equation}
+{i\over2}
\sigma^{\,\alpha}{}_{\delta\,,\,\phi}\,
\sigma^{\,\phi}_{\;\;\;\beta\,,\,\alpha}
\sigma^{\gamma\beta}
=-{1\over2}
\sigma^{\,\phi\gamma}_{\;\;\;\;\,,\,\alpha}\,
\sigma^{\,\alpha}_{\;\;\;\delta\,,\,\phi}
\label{odd}
\end{equation}
which sums with (B).(E'). Analogous 
contributions arise from (C).(F') and
(C').(I).
The crossed term
(B).(F'), (B').(I),
(C).(E') and (C').(H), 
each give;
\begin{equation}
{i\over2}
\sigma^{\,\alpha}_{\;\;\;\beta\,,\,\phi}\,
\sigma_{\alpha\delta\,,}{}^{\phi}
\sigma^{\gamma\beta}
\equiv
+{1\over2}\partial^{\phi}A^{\gamma}
\partial_{\phi}A_{\delta}
\label{Aterm2}
\end{equation}
For similar reasons the product
(A).(E') gives
\[-1/2\,\partial^{\gamma}A^{\alpha}
\partial_{\alpha}A_{\delta}\]
and similarly for (A).(F'),
(B').(G) and (C').(G).
Products (B).(D'), (C).(D'),
(A').(H) and (A').(I) are easily
evaluated and each gives
$-{1\over2}\partial^{\phi}A^{\gamma}
\partial_{\delta}A_{\phi}$.
To translate
(\ref{odd})
the $\alpha$ contraction on the indices
delivers a $+i\partial_{\,\delta}$ and the
$\phi$ contraction a $-i\partial^{\,\gamma}$
; the -i sign because with relabelling
it can be seen that the two matrices 
\[
\sigma^{\phi\gamma}_{\;\;\;\;,\alpha}\;
\sigma^{\alpha}_{\;\;\;\delta\,,\,\phi}
\]
anti-commute. Hence we obtain; 
\begin{equation}
-{1\over2}
\sigma^{\,\phi\gamma}_{\;\;\;\;\,,\,\alpha}\,
\sigma^{\,\alpha}_{\;\;\;\delta\,,\,\phi}
\equiv
+{1\over2}\partial^{\gamma}A^{\phi}
\partial_{\delta}A_{\phi}
\label{acontraction}
\end{equation}

Summing the non-zero components of the
tensor part we have;
\begin{eqnarray}
+2i
R^{\alpha}_{\alpha\delta\beta}\,
\sigma^{\gamma\beta}
=\,&\,&
-2R^{\gamma}{}_{\delta}=
+2R_{\delta}{}^{\gamma}
\nonumber\\
{\equiv}\,
&-&2\partial^{\gamma}A^{\alpha}
\partial_{\alpha}A_{\delta}
+2\partial^{\alpha}A^{\gamma}
\partial_{\alpha}A_{\delta}
\nonumber\\
&+&2\partial^{\gamma}A^{\alpha}
\partial_{\delta}A_{\alpha}
-2\partial^{\alpha}A^{\gamma}
\partial_{\delta}A_{\alpha}
\nonumber\\
{\equiv}\,&\,&2F^{\gamma\,\alpha}
F_{\delta\,\alpha}
\label{F}
\end{eqnarray}
Raising indices with the symmetric
part of the metric we finally
obtain the traceless electro-magnetic
stress-energy tensor;
\begin{equation}
-{1\over{\kappa^2}}R^{\delta\,\gamma}
+{1\over2{\kappa^2}}\eta^{\gamma\delta}
R=F^{\gamma\,\alpha}
F_{\alpha}^{\;\;\delta}+{1\over4}
\eta^{\gamma\,\delta}
F^{\mu\,\nu}
F_{\mu\,\nu}
\label{set}
\end{equation}

In forming equation (\ref{set}) I have replaced
the equivalence relation ($\equiv$) by an = sign
and a dimensional constant $\kappa^{-2}$
(the gravitational coupling constant).
 This is discussed 
in section IX.

The derivation of the traceless gauge-invariant
free-field stress-energy tensor equated to the
Einstein-like equation is something of a
mathematical miracle. There must be exactly the
right number and type  of non-zero pieces to construct the 
tensor and the factor of 2 difference between the
scalar R and the tensor $R^{\delta\gamma}$ on the
L.H.S. of eq.(\ref{set}) gets translated into an
effective factor of 4 difference on the R.H.S. 
only because of the spinorial representation
used and the translation procedure. This is a 
actually a  non-trivial result. I suspect it is 
the only way a traceless gauge-invariant F.F.
tensor can be extracted from a standard
Larangian.

The issue of the anti-symmetry in
$\delta$ and $\gamma$
of (\ref{F}) is discussed below.

\section{discussion of homothetic curvature}

The above derivation effectively eliminates the
symmetric part of the metric.
Applying the anti-symmetry constraint $\mu\neq\nu$
for a purely anti-symmetric metric;
\begin{equation}
\bar{g}_{A}^{\mu\nu}=
-i\sigma^{\mu\nu}={1\over4}
[\gamma^{\mu},\gamma^{\nu}]_{-}
={1\over2}\gamma^{\mu}\gamma^{\nu}
_{\;(\mu\neq\nu)}
\label{vector}
\end{equation}
so, using (\ref{P});
\begin{eqnarray}
&\sigma&^{\nu\mu\,,\,\delta}
\sigma_{\nu\delta\,,\,\mu}
={1\over4}\partial^{\delta}\left(
\gamma^{\mu}\gamma^{\nu}\right)
\partial_{\mu}\left(
\gamma_{\nu}\gamma_{\delta}\right)
\nonumber\\
&\approx&
{1\over4}\left(\partial^{\delta}
\gamma^{\mu}\right)\gamma^{\nu}\gamma_{\nu}
\left(\partial_{\mu}\
\gamma_{\delta}\right)
=
\left(
\partial^{\delta}
\gamma^{\mu}
\right)
\left(
\partial_{\mu}
\gamma_{\delta}
\right)
\label{cont}
\end{eqnarray}
which identifies the $A^{\mu}$ field as a 
$\gamma^{\mu}$ 4x16 matrix
$
|P|\gamma^{\mu}%
\equiv
A^{\mu}
$
transforming as a 
{\it vector} under the 16x16 anti-symmetric
metric (\ref{vector});
{\it with respect to commuting
co-ordinates} (note; 
use of commuting co-ordinates
implicit
in the derivation of results
- note also that (\ref{P}) implies that
the A field is only defined up to a local phase).
Normally an infinitessimal rotation is given
by
\[ \delta x^{i}=\epsilon^{ij}\eta_{jk}x^{k}
=\epsilon^{i}_{\,\,\,k}x^{k}\]
where $\epsilon^{ij}$ is antisymmetric.
Now $R^{\gamma}_{\;\;\delta}$ in 
(\ref{F}) is anti-symmetric
but under $g^{A}$ an infinitesimal rotation
is given by; $\delta x^{i}=s^{ik}g^{A}_{kj}x^{j}$
where $s^i_{\,j}$ is symmetric thus variation of the
Lagrangian \cite{three} (for generic field $\phi\;$)
will give; 
\begin{eqnarray}
0&=&
\nonumber\\
&s&_{\mu\nu}\partial_{\rho}
[
\frac{\delta\cal{L}}{\delta\partial_{\rho}}
(\partial^{\mu}\phi x^{\nu}+\partial^{\nu}
\phi 
x^{\mu})
-
g^{\rho\nu}x^{\mu}\cal{L}-
\mbox{\it g}^{\,\rho\mu}
\mbox{\it x}^{\nu}\cal{L}
\nonumber
]
\end{eqnarray}
with the divergence of the conserved current;
\[
\partial_{\rho}
\cal{M}^{\rho\,,\,\mu\nu}=T^{\mu\nu}
+T^{\nu\mu}
\]
which is zero if the stress-energy tensor
$T^{\mu\nu}$ is {\it anti-symmetric}
under $g^{A}$ (in other words, in the framework
of an anti-symmetric metric the stress-energy
tensor must be {\it anti-symmetric} to obtain
conservation of angular momentum - this is the
opposite to the situation with a purely symmetric
metric where the stress-energy tensor must be
{\it symmetric} to conserve angular momentum). 

Effectively
we have a choice of description;
(1) we can describe the A field as a 
conventional vector
with symmetric metric in commuting co-ordinates,
or (2) as a `$\gamma$' vector with anti-symmetric metric
in commuting co-ordinates.
We know from (\ref{vector})
that $A^{\mu}$ must transform as a
4 vector under the
space-time metric.  
Raising indices in
(\ref{set}) with the
diagonal part of (\ref{metric}) implies we
revert to description (1) instead of
(2) where the $A^{\mu}$ is no-longer
a $\gamma^{\mu}$ vector but a simple 4-vector
transforming under symmetric metric
and $T^{\gamma\delta}$
is instead symmetric because angular momentum
conservation must be present regardless
of the choice of description. 
However, this of course 
means that we must equivalently substitute
 a {\it symmetric}
$R^{\gamma\delta}$ in eq (\ref{set}) for 
the anti-symmetric
value that arises in eq (\ref{F}). This
 relates back to the 
A-field definitions like eq (\ref{acontraction})
 the notation 
of which is appropriate for a symmetric term.
 It is the L.H.S.
of eq (\ref{acontraction}), and analogous
 contributions,
which should properly be summed to form the
 antisymmetric object
$R^{\gamma}_{\;\;\delta}$
 in eq (\ref{F})
; the conventional A-field definition (in commuting
 co-ordinates) 
is only
appropriate when we translate to the symmetric objects
 (i.e. eq(\ref{set})). I have introduced the A-field
notation ( eq(\ref{Adefinition}), eq 
(\ref{acontraction}) etc)
early as this facilitates comprehension and also
 demonstrates that
there is a consistent mathematical method for
 performing the 
translation. It must be noted however that
 there is always
an inherent choice of sign on the tensor part when we 
perform a translation between an anti-symmetric and a
symmetric object; this is the price we pay
for working in a spinorial representation
against an anti-symmetric metric which becomes
a representation {\it{up to a sign}}.
 For example eq (\ref{delta}) can
be rewritten as;$\left({1\over2}
\eta^{\phi\delta}R_A+R^{\phi\delta}
\right)_{;\,\delta}=0$ where the 
tensor part now has opposite
sign. It can however be argued that
the same traceless stress-energy 
tensor will result since we can choose
a sign from the residual phase factor
from the index raising operation in eq(\ref{array2})
(the phase can be made to vanish for the scalar $R_A$).

Lastly in this section note that the 
Lagrangian for the free-field is now
given by the scalar curvature;
\begin{equation}
{\cal{L}}= {1\over\kappa^2}
{\sqrt{|g|}}{\lambda}R=-{1\over4}
F_{\alpha\beta}
F^{\alpha\beta}
\label{lagrangian}
\end{equation}
 where  $\lambda$ is a normalisation 
constant. Because
the representation effectively normalises
the A field the norm of g
is a constant and can be absorbed into
the $\kappa$.
Just how the $\kappa^2$ gravitational
constant is absorbed into the free-field
is discussed later in the text.

\section{Rotation curvature and source
terms}

Re writing eq.(\ref{dot}) for conventional 
curvature we have;
\begin{equation}
R^\alpha_{\gamma\beta\alpha}=\partial
_{\beta}\,\Gamma^{\alpha}_{\gamma\alpha}
-\partial_{\alpha}\,\Gamma^\alpha_{
\gamma\beta}
-\Gamma^{\alpha}_{\phi\beta}\,\Gamma^
{\phi}_{\gamma\alpha}+
\Gamma^{\alpha}_{\phi\alpha}\,
\Gamma^{\phi}_{\gamma\beta}
\label{dot2}
\end{equation}
 In General Relativity the symmetric metric
feeds into the rotation curvature 
viz the symmetric connection and 
defines the stress-energy tensor viz
Einstein's equation. It is relatively easy
to prove that the rotation curvature is 
zero for the component of metric (\ref{metric})
that is on the light cone (i.e. the antisymmetric
part of the metric). We identify the rotation 
curvature with material sources; i.e. particles
with mass and these sources should  be identified
with the symmetric part of the metric. The phase
factor identified with the $ds^2=0$ part of the
metric (the $A_\mu$ field)
contains an implicit factor of $\hbar=1$ and thus
imply a `waviness' to space-time structure
at the quantum scale. It is this wave-like
structure of small-scale space-time that has
replaced the fifth dimension of Kaluza-Kline
 theory; the space-time structure itself has
been given the properties of the harmonic
oscillator. 

Thus in order to obtain particle sources for
the theory we must now modify the small-scale
structure of space-time for the symmetric part
of the metric. The appropriate phase factor
will now be based on particle momentum 
and the metric takes the form ($\hbar=c=1$);
\begin{equation}
g_{\mu\nu}=
e^{ip{\cdot}x}I_4.\eta_{\mu\nu}
+ie^{ik{\cdot}x}\sigma_{\mu\nu}
\label{sourcemetric}
\end{equation}
where $p^{\alpha}$ is the source four-momentum
and $k^{\alpha}$ is the photon four-momentum.

The first two components of the expansion of
the rotation curvature (R.H.S. \ref{dot2}) are zero
since;
\[
\partial_{\gamma}
\left(
\partial_{\beta}e^{ip{\cdot}x}
\right)
e^{-ip{\cdot}x}
=ip_{\beta}\partial_{\gamma}
\left(e^{ip{\cdot}x}e^{-ip{\cdot}x}
\right)
=o
\]
so there is no interference
with gravitation at the level of 
the E.M. sources and we have;

\begin{equation}
R^\alpha_{\gamma\beta\alpha}=
-\Gamma^{\alpha}_{\phi\beta}\,\Gamma^
{\phi}_{\gamma\alpha}+
\Gamma^{\alpha}_{\phi\alpha}\,
\Gamma^{\phi}_{\gamma\beta}
\end{equation}

For the source the metric is symmetric and
the connection takes the usual symmetric 
form. We may
take it as identical to (\ref{zot})
with the anti-symmetric part omitted
and thus we obtain (for notational
convenience dropping the $I_4$
and absorbing the phase-factor 
into the definition of $\eta$ in 
an analogous manner as was done with the
$\sigma$ matrices);

\begin{eqnarray}
4R^\alpha_{\gamma\beta\alpha}&=&
-
\left(\eta_{\phi\;\;,\beta}^{\,\,\,\alpha}+
\eta^{\alpha}_{\;\;\beta\,,\,\phi}
-\eta_{\phi\beta\,,}^{\;\;\;\;\;\alpha}
\right)e^{-ip{\cdot}x}.
\nonumber\\
&\,&\;\;\;\;\;\;\;
\left(\eta_{\gamma\;\;,\alpha}^{\,\,\,\phi}+
\eta^{\phi}_{\;\;\alpha\,,\,\gamma}
-\eta_{\gamma\alpha\,,}^{\;\;\;\;\;\phi}
\right)e^{-ip{\cdot}x}
\nonumber\\
&+&
\left(\eta_{\phi\;\;,\alpha}^{\,\,\,\alpha}+
\eta^{\alpha}_{\;\;\alpha\,,\,\phi}
-\eta_{\phi\alpha\,,}^{\;\;\;\;\;\alpha}
\right)e^{-ip{\cdot}x}.
\nonumber\\
&\,&\;\;\;\;\;\;\;\;
\left(\eta_{\gamma\;\;,\beta}^{\,\,\,\phi}+
\eta^{\phi}_{\;\;\beta\,,\,\gamma}
-\eta_{\gamma\beta\,,}^{\;\;\;\;\;\phi}
\right)e^{-ip{\cdot}x}
\nonumber\\
&=&
+2\left(\partial_{\gamma}e^{ip{\cdot}x}\right)
e^{-ip{\cdot}x}
\left(\partial_{\beta}e^{ip{\cdot}x}\right)
e^{-ip{\cdot}x}
\nonumber\\
&-&2\eta_{\gamma\beta}
\left(\partial_{\phi}e^{ip{\cdot}x}
\right)e^{-ip{\cdot}x}
\left(
\partial^{\phi}e^{ip{\cdot}x}\right)
e^{-ip{\cdot}x}
\end{eqnarray}
from which we obtain;
\begin{equation}
R^{\alpha}_{\gamma\beta\alpha}
=
-{1\over2}p_{\gamma}p_{\beta}
+{{m_{o}^2}\over2}\eta_{\gamma\beta}
\end{equation}
where $m_{o}^2$ is the
square of the rest mass and
\begin{equation}
R=R^{\alpha}_{\gamma\beta\alpha}
\eta^{\beta\gamma}
=
+{3\over2}m^2_{o}
\label{nophase}
\end{equation}
so
\begin{equation}
-R^{\alpha}_{\gamma\beta\alpha}
+{1\over2}
\eta_{\gamma\beta}R
=
{m^2_{o}\over2}\left(
U_{\gamma}U_{\beta}+{1\over2}\eta_{\gamma\beta}
\right)
\label{particlestressenergytensor}
\end{equation}
where $U_{\gamma}={{dx_{\gamma}^p}
\over{d\tau}}$ where $\tau$ is the
proper time and p denotes the particle
position. I have supressed the
phase factors associated
with the symmetric
metric contraction in
eq(\ref{nophase}) because what is being
performed here is a translation to a 
classical description of a point
particle. We will eventually add a
factor of dimension $l^3$ 
to the symmetric part of metric
(\ref{sourcemetric}) in order
to create a Lagrangian density of the 
appropriate dimension. With translation a
factor of $l^{-3}$ will appear in the 
fields. In anticipation
of this we add a factor of dimension $l^{-3}$ to
obtain a translation to a classical
particle description with  position $x(\tau)$ 
and use;
\begin{eqnarray}
d^3(x)&=&{\int}d\tau\;\delta\left(
x^o-x^o_p\tiny{(\tau)}\right)
\delta^3\left(x^i-x^i_p\tiny{(\tau)}\right)
\nonumber\\
&=&
{d\tau\over{dx^o}}\delta^3(x^i-x^i_p
\tiny{(\tau)})
\nonumber\\
\label{61}
\end{eqnarray}
so that eq(\ref{particlestressenergytensor}) finally
gives (dropping the factor of 1/2 which is analogous
to the zero-point energy of an harmonic oscillator);
\begin{equation}
{m_o\over2}{\lambda}T_{\gamma\beta}
=
{m^2_{o}\over2}
U_{\gamma}U_{\beta}
{d\tau\over{dx^o}}\delta^3(x^i-x^i_p
\tiny{(\tau)})
\end{equation}
which is the correct form for the classical
stress-energy tensor for a point-particle with
unit charge \cite{CFT}. The three-dimensional
delta function has been substituted for the
fields in the classical description (c.f eq(\ref{st})).
(The delta function, in the classical limit that the 
space-time spread for the particle approaches
a point, behaves as the inverse of the irreducible
metric - see eqs (\ref{I.I.}),(\ref{I.A.})
and (\ref{I.V.}) - thus the use of the delta function is
only valid in the `classical limit' and not
in a quantum description in which case
the irreducible metric can not be treated as
the inverse of a delta function). 
 Lambda is a constant to be 
determined.
Note that I have used the same sign
convention for the particle stress-energy
tensor that I employed for the free-field
stress-energy tensor for consistency
(see the section titled Homothetic Curvature).
 The origin of the
zero-point additional energy is analogous to
the non-vanishing of the zero-point energy of
a simple harmonic oscillator that is
see in quantum physics. It
is an indication that the transition to
the point-particle description is not
entirely appropriate.  Note also
that, due to Einstein's equation, the
covariant derivative of the 
particle stress-energy tensor
vanishes in the absence of the free-field. 

\section{The concept of irreducibility}

In four dimensions the Lagrangian density
must have dimension $l^{-4}$. Formally the
metric must be dimensionless. This immediately
leads to a problem with the theory
presented above as follows.
The Lagrangian density ${\cal{L}}=-{1\over4}F
_{\alpha\beta}F^{\alpha\beta}$ has dimension
$L^{-4}$ because the $A^\alpha$  field is given
dimension $l^{-1}$ and each derivative contributes
an $l^{-1}$.

The contracted curvature tensor (whether homothetic
or rotation), when derived from a dimensionless
metric, thus has dimension $l^{-2}$. It is this
fact that makes the coupling constant of the 
graviational field $l^{-2}$ and renders quantum
gravity non-renormalisable. 

Thus it appears that in performing the translation
between symmetric and anti-symmetric representations
of the electro-magnetic field we must also
introduce a dimensional transmutation in
order to give the free-field Lagrangian the 
appropriate dimension.

Ultimately this is the crux of the 
problem of unifying electro-magentism and
gravitation and also the central issue
causing the difficulty quantising gravitation.
Very much in the spirit of H Weyl's ideas, I
want now to explore a possible solution to
this problem that centres about the issue of
scale-transformations.  The following
is a sketch, not entirely rigorous, of
the central ideas involved.

Einstein hints at the problem in his last
published paper
\cite{UFT}  when he discusses the 
obvious difference between the inherent
discontinuity of quantum objects and 
the continuum of space-time; an apparent
schitzophrenia that has no deeper physical 
explanation in current theory.

Let us firstly assume that discontinuity
is the fundamental element of phsical
structure and that the continuum is built
up from a more fundamental element of 
structure that is ultimately completely
discontinuous. This would imply that both
matter and space-time are built from the
same basic `stuff'. (A strong empirical
hint that this must be the case is 
seen with phenomena such as
 creation of particle pairs
from the vacuum in HEP). The most basic 
element of structure that could be
postulated seems to me to be something
like an `on-off' or (0,1) duality. A
plausable associated metric would be;

\begin{equation}
d(a,b)=|{\epsilon}(a,b)|
\label{I.I.}
\end{equation}

The meaning of eq.(\ref{I.I.})
is that the distance between points labelled
a and b is the absolute value unity (i.e. 1)
if $a{\neq}b$ and zero if $a=b$. This is, of 
course, regardless of where the points a and
b happen to be located. Indeed, according to this
metric it makes no sense to talk about where the
points are; only that they are separate or
distinguishable. No `background' space-time
as such exists according to this metric; we want
to {\it{build}} a four-dimensional space-time
out of this metric. We postulate the following 
algebra for the metric;

\begin{equation}
 |\epsilon(a,b)|.|\epsilon(b,c)| = |\epsilon(a,c)|
\end{equation}
so that the product of two objects of 
dimension $l^1$ is not $l^2$ but $l^1$. I call such an
object an {\it{irreducible interval}} and its dimensionality
is also set irreducibly at unity;
dimensionality is thus in some
sense quantised in this scheme. The {\it{number}} one is
defined as a {\it{counting}} of the existence of the 
interval from one end to the other. Iterated countings
still only define the number one. The number zero may
be thought of as the non-existence of the interval or the
point upon which counting is initiated.

Iterated counting may be symbolised as;
\begin{equation}
|\epsilon|^n=1\;\;\;({\forall}n\neq\aleph_0)
\end{equation}

The object
$|\epsilon|^{\aleph_0}$ 
with transfinite (completed infinite) index
is not definable in a singular irreducible
dimension. We assume it defines a two dimensional
space bounded by irreducible intervals.
Such a space must contain at least three points
on its boundary. 
Its associated metric is written as;
\begin{eqnarray}
d^2(a,b,c)=|\epsilon^2(a,b,c)|=|\epsilon^2|
\nonumber\\
|\epsilon|^{\aleph_0}=|\epsilon^2|
\label{I.A.}
\end{eqnarray}
The `area' bounded by the irreducible intervals
and defined by metric (\ref{I.A.}) I will
call an `irreducible area' or I.A. Its 
cardinality is that of the counting numbers
$\aleph_0$
(i.e. the field of rational numbers)
 {\it{not}} that of the continuum.
(By contrast the cardinality of the irreducible 
interval is strictly finite).
It is this kind of object that I want to assume
forms the superstructure of the photon. On the light 
cone we assume it doesn't define a space with the 
property of the real continuum. To get a continuum
we must assume the continuum hypothesis (i.e.
that the next highest transfinite cardinal above
$\aleph_0$ is c the cardinality of the continuum),
and that propagation of the photon with respect to
all and any  observers generates such an equivalent space.
The metric may be written;
\begin{eqnarray}
d^3(a,b,c,d)=|\epsilon^3(a,b,c,d)|=|\epsilon^3|
\nonumber\\
|\epsilon^2|^{\aleph_0}=|\epsilon^3|
\label{I.V.}
\end{eqnarray}
(The last equation
is analogous to the equation $2^{\aleph_0}=c$).
Metric (\ref{I.V.}) describes {\it{irreducible 
volumes}} (I.V.'s) the contained space of which is assumed
to have the mathematical property of the continuum
but no contained fourth dimension (no time). Such a
space provides a candidate for both quantum objects
with mass and propagating photons; of course for the 
latter we must add time as the dynamical factor 
generating the volume if we assume that photons
are propagating I.A.'s. with respect to objects with
mass. On the boundary of massive objects we 
will expect to find I.A.'s and thus an associated
massless field.

Of course with this kind of senario the time
dimension itself is not really geometrically
defined; it is an assumed added parameter. It is 
possible to extend the geometric/mathematical
analogy to postulate a more geometric origin
for time but here we will assume that the
addition of time does not alter the
cardinality; space-time has the same cardinality
as 3-space which is that of the  continuum.

Note that, even though a timeless 3-space
defined by metric (\ref{I.V.}) is a continuum
it is {\it{irreducible}} in the sense that
it cannot be subdivided because to do so would violate
the irreducibility of the bounding intervals (
or equivalently the bounding areas) upon which the
hierarchy of structure is built; it is in this
sense  quantised
irreducibly and immortal. Dynamics can only
occur on the boundary of the object; never in its
interior.

It is possible to postulate that the 
 physical manifestation of metrics (\ref{I.I.}),
(\ref{I.A.}) and (\ref{I.V.}) is
local gauge invariance. To see how this idea
works geometrically consider three
points selected at random 
on a circle consisting of 
a real continuum of points;

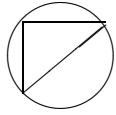
\setlength{\unitlength}{1mm}
\begin{picture}
(30,30)(-25,0)
\put(15,15){\circle{20}}
\put(10,10){\line(6,5){11}}
\put(10,10){\line(0,0){9.5}}
\put(10,19.5){\line(4,0){11}}
\end{picture}

Now, we know form the work of G. Cantor
that for a continuum of points a 1:1
mapping can be defined from the points on 
any finite length line segment onto any other
line segment of arbitrary length.
 Thus for the continuum of 
points on the circle we can define a
1:1 and onto mapping of the circle onto
itself which does not leave the triangle
invariant; the three points defining
the triangle can be shifted around the
circle by such a transformation. This is
another way of saying that the continuum
can be compressed or stretched to an
arbitrary degree without the structure of
the continuum itself varying. Such a mapping is an exact
analogue of a local U1 gauge transformation
on the circle. However, under metric (\ref{I.I.}),
and indeed only under metric (\ref{I.I.}),
the triangle itself may be regarded as invariant
since the angles subtended by the sides of the triangle
are not defined under such a metric since each edge
of the triangle always has unit length. Such a 
triangle is `irreducible' under a local gauge 
transformation. Alternatively we may view
the concept of the combination of an
irreducible geometry  embedded in
the structure of the continuum as a
deeper explanation of the origin of
local gauge invariance itself; i.e. that the 
embedding of absolute discontinuity into
the continuum gives rise to local gauge
invariance. I have in mind here the basic 
foundation of quantum objects embedded in
the continuum of space-time; or, 
alternatively in the language of the
geometry presented above, of quantum 
objects actually {\it{generating}} the
space-time continuum. Such an embedding
is a fundamental union of the discrete and the
continuous.

We now postulate that a photon literally
has intrinsic geometric structure built up
from irreducible intervals which have geometric
and physical definition only on the light cone
itself (a triangular geometry, for example, 
might be candidate) or
more particularly as some
form of irreducible geometry
defined in a two-dimensional plane
orthogonal to the direction of motion
of the photon and propagating
at the speed of light. The geometric irreducibility,
which is inherently non-local,
itself is unobservable; we see its physical 
manifestation {\it{indirectly}} through the
unobservability of local gauge; i.e. local gauge 
invariance of electro-magnetism. (Of course the same
must apply to the {\it{boundary}} of a three-dimensional
object defined by metric (\ref{I.V.}); such
a geometry is assumed to be a massive fermion
quantum object and the boundary its associated
electro-magnetic and gravitational fields; there 
must be 
implications here for the theory of neutrinos
but I will not discuss this issue in this paper).

We can now reinterpret the translation process
for the free-field electro-magnetic stress-energy
tensor as follows. The anti-symmetric part of 
metric (\ref{metric}) is an irreducible 
metric on the light cone; this means that it
does not hold in any observer's frame. 
Each $\sigma_{\alpha\beta}$ term,
which ultimately will contribute one
$A_\alpha$ or  $A_\beta$ 
term,
is assumed to have dimension $l^{-2}$ and,
in addition to a dimensionless
phase factor $e^{{\pm}ik{\cdot}x}$,
contains an intrinsic product of an
I.A.  to make the whole object (c.f. eq(\ref{|P|}))
\begin{equation}
g^A_{\alpha\beta}=
|\epsilon^2{\tiny{(\alpha,\beta)}}|.
e^{{\pm}ik{\cdot}x}.\sigma_{\alpha\beta}
=
|P|\sigma_{\alpha\beta}\equiv
\sigma_{\alpha\beta}
\label{irredmetric}
\end{equation}
dimensionless. (The I.A. here is rather
like a dimensional polarisation tensor; because of the
peculiar algebra of these metrics we can
still use the $g_A$ to raise and lower
indices).

Now the dimension $|\epsilon|$ is `infinitely smaller'
than the dimension $|\epsilon^2|$. In anticipation
of imposing a scale on the irreducible metric as
a part of the translation procedure let us 
define the irreducible area viz a term $d|\epsilon^2|$;
\[
|\epsilon^2|=
\int_{-\infty}^{+\infty}|\epsilon|.\;d|\epsilon^2|
\]
where $d|\epsilon^2|$ is the (infinitessimal but 
denumerable) increment
in area in a direction orthogonal to $|\epsilon|$.
The integration is carried over all space.
Also we have;
\[
|\epsilon^3|=
\int_{-\infty}^{+\infty}|\epsilon^2|.\;d|\epsilon^3|
\]
With translation to an observer frame the I.A.
ceases to exist (we
 must assume that it becomes absorbed into the
structure of the continuum
\cite{GF}) and the continuum 
has its dimension boosted from two to $3+1$
dimensions. The idea is to regard the metric
$|\epsilon|$ as related to the graviton
(this metric must be spin-2 since the 
generator of the discrete
associated group $S_2$, the permutation group
of two objects, will not change sign with a 
rotation by $\pi$), the metric $|\epsilon^2|$
as related to the photon (the three-point 
discrete group C3v of the triangle changes
sign with rotation by $\pi$ of one of
its generator axes) and the metric
$|\epsilon^3|$ as the metric of a 
massive spinor (viz-a-viz the 4-point
discrete spinor group Td). The `smallness'
of $|\epsilon|$ may then be expressed in 
translation to the observer frame viz a 
Taylor expansion;
\begin{equation}
|\epsilon^2|\approx
(\partial_{\mu}|\epsilon^2|).\kappa
\approx\kappa^2
\label{translation}
\end{equation}
Which implies that in translation to the
observer frame the metric $|\epsilon|$ and
the $d|\epsilon^2|$ are
`small' in relation to the electro-magnetic
irreducible metric to the 
order of the gravitational constant. Notice that with 
translation (\ref{translation})
the metric equation $|\epsilon|.|\epsilon|
=|\epsilon|$ 
no longer holds because a scale has been set.

Similarly, we assume that the volume increment
 $d|\epsilon^3|$ 
is `small' with respect
to the volume metric $|\epsilon^3|$ to the
order of the fine structure constant in relation
to the mass-scale $m_o$ of the object defined by the 
three-dimensional metric. Thus with translation we set;
\begin{equation}
|\epsilon^3|\approx
|\epsilon^2|.{\alpha\over{m_o}}
\approx{\kappa^2\alpha\over{m_o}}
\label{3trans}
\end{equation}

Lastly we must impose ananalogous set of conditions
on the symmetric part of the metric;
\begin{equation}
g^S_{\alpha\beta}=
|\epsilon_i^3|e^{ip_i{\cdot}x}\eta_{\alpha\beta}
\end{equation}
where $\eta_{\alpha\beta}$ now contains a 
product of two fields each
of dimension $l^{-{3\over2}}$ i.e. spinor fields,
and the irreducible  volume metric of
dimension $l^3$ appears.
The i here labels particle types.
We must assume that with a translation procedure
from a symmetric to an anti-symmetric representation
(in some sense counter-balancing the translation
of the boson field $A_\mu$ from an anti-symmetric to a
symmetric representation)
the $\eta$ field translates
into a spinorial  representation
(using (\ref{3trans})) viz the ansatz;
\begin{equation}
|\epsilon_i^3|e^{ip_i{\cdot}x}
\eta_{\alpha\beta}
\equiv
{\kappa^2\alpha\over{m_o}}\;
\overline{\Psi}_i{\Psi_i}\gamma_{\alpha}\gamma_{\beta}
\label{st}
\end{equation}
of the spinor $\Psi_i$.
Note that the R.H.S.of eq.(\ref{st}) is
antisymmetric in $\alpha$ and $\beta$ so this
involves a translation between symmetric and
anti-symmetric representations (i.e. we don't
equate both sides to zero!).
Whether or not the Dirac Lagrangian can
be extracted from this form of
translation remains to be
seen.
The special algebra of these irreducible
metrics, as
before, allows the use of the total metric
to raise and lower indices. The
irreducible metric $|\epsilon^3_i|$  behaves
algebraically like a dimensionless quantity
prior to translation. 
Our Lagrangian  reads;  
\begin{equation}
{\cal{L}}
=
{\kappa^{-2}}{\sqrt{|g|}}R
\label{Lag}
\end{equation}

\section{discussion; supersymmetry or 
superslimmetry?}

In this paper only the photon has
been given `dual' representations both
as spinor and vector. (Equation (\ref{st})
is just a speculation for further work).
The `operator'
which interconverts the photon representations is
the  procedure that converts 
homothetic curvature due to
torsion  into rotation curvature. What does this
mean geometrically?

Consider again the triangle embedded in
the circle pictured previously. Torsion breaks
parallelograms (or equivalently triangles)
but under the irreducible metric
the triangle does not break when subjected to
torsion. In fact it's `unbreakability' under
the torsion induced homothetic curvature
is nothing other that an expression of
 U1 local-gauge invariance
as was previously demonstrated. But this is
a compact rotation symmetry! Thus we see that
the interconversion of bosonic and fermionic
representations is bridging compact and non-compact
groups because the torsion is the generator of
translations. This enables us to have 
a more fundamental physical reason for the 
occurrence of local gauge invariance in
nature; invariance of the structure of the
continuum to arbitrary deformations.
 It is the
structure of the continuum which is truely
fundamental; the matter fields and the 
forces between them appear as the superstructure
keeping the continuum continuous.

Looking at the invariance of the irreducible  geometry
used to describe the photon
under torsion induced translation is equivalent,
at least from the geometric point of view,
to `dressing' the photon with its own
gravitational self-interaction. To see
this note that, since torsion breaks parallelograms
if the triangle were defined by
 ordinary geometry it  would break
if the intervals defining it were not irreducible. 
Consider the simplest break; a rupture at one
of the vertices of the triangle. The result
will be an object with four vertices. The
extra interval, under the assumptions presented,
is  the geometry of a graviton. The restoration
of the geometry of the triangle would then correspond
to the resorption of the emitted graviton. In
this manner irreducibility of the
geometry of the triangle - that is,
its invariance under torsion induced
translations - is equivalent to `dressing'
the photon with the gauge field of translations;
its own gravitational self-interaction. 
Eq.(\ref{translation}) gives us
a scale for the interaction; each photon is
dressed by gravitons of the order of the
Planck length. Thus the `breaks' induced by 
torsion are extremely small scale.

A similar interpretation can be given to
eq.(\ref{3trans}); irreducibility equals
local gauge invariance and here the global
U1 phase of the photon is the gauge group
restoring invariance of the phase of the 
source of the electro-magnetic field. The
irreducible metric is implicitly including
the gravitational and electromagnetic 
self-interactions of the the source; the
source is `dressed' by its own fields.
 
It seems to me that we have the following
senario. We have dual descriptions of 
electromagnetism. Interpreted in an observer
frame torsion and its induced homothetic
curvature is unobservable; it could only
ever appear as a non-propagating contact
interaction. However, to
an observer `travelling on a ray of light'
(which, for the sake of explanatory convenience,
we shall admit)
the torsion and homothetic curvature
would appear real and the observers 
world would be a strange place where photons 
behave as spinors and only the anti-symmetric
part of the photon's stress-energy tensor is a 
conserved quantity. To our observer riding
on a photon the photon obeys a Pauli exclusion
principle; which is to say in a space where
$ds^2=0$ our observer is in no position to 
`see' any other photon other than the one he
or she is unfortunate enough to be ensconced
with.

Then there is the other description of 
electromagnetism with which we are more
familiar. In it the photons are bosons
and the conserved stress-energy tensor is 
of course symmetric. In this frame the
homothetic curvature to our observer
perched on a photon appears
to
the more familiar observer on Earth 
 as rotation
of space-time in the local vicinity of 
each individual photon 
when  we call it
the wave nature of light.

It is in this manner that 
the C-M theorem may be overcome; not
by supersymmetry but with a slimmer
menagerie of fundamental objects -
which is of course desirable -
with each particle providing its own 
`superpartner' of which the photon,
in this case, may provide the 
prototype example. This seems to me  
more natural than conventional 
supersymmetry given that it generates
local gauge invariance rather than 
assuming it and does not generate 
unobserved objects.

Thus  
 unification
in four dimensions is not yet a 
closed subject and hopefully this
paper has stimulated some interest
in it. In particular
obtaining a gauge-invariant traceless
stress-energy tensor for the free-field
is quite a non-trivial result peculiar
to the mathematical structure I
have presented. Note that 
 exactly the right components must
be present in the expansion of the 
curvature tensor for the mathematics to work.
Variation of the Lagrangian (\ref{Lag})
now leads to the Einstein equation on the L.H.S.
and the sum of the free-field and particle
stress-energy tensors on the R.H.S. and both
gravitation and electro-magnetism are
accommodated in the single equation. The 
unification of the gauge couplings is 
speculative but is assumed to be linked 
to the structure of the continuum beyond
the Planck scale. 
Using irreducible metrics means that we really
must go beyond the conventional conception of
space-time. Instead of a fifth dimension to
define electric charge, particles now appear
rather like non-local bubbles in the vacuum inside which time
is absent.  The closer we look
at the bubble the smaller it gets
(I have in mind here electrons and quarks).
The quantisation of charge
must now be related to the 
topology of the boundary of this space.
The decomposition of the vacuum is more
severe on the light cone where the structure of the
continuum itself is actually altered. 

It would be appropriate to summarise what
has been done in order to get the 
mathematical content in perspective.
Firstly the metric structure of space-time
has been generalised;

 1. to include a $U(1)$ 
phase factor `on the light-cone' the generator
of which is the photon momentum. The sigma
matrices in some sense `carry' the representation
$e^{{\pm}ik{\cdot}x}$
on the light cone. The phase factor itself means
that, in the presence of the photon, space
on the light cone has an intrinsic wave-structure.
The associated torsion is generating, viz the
homothetic curvature, the corresponding free-field
stress-energy tensor. 
This, however, is not really
a true Riemann-Cartan geometry because
the torsion on the light
cone translates to non-torsional
rotation curvature in the
`observer frame'. The `frame' in which this 
torsion is defined is `on the light cone' i.e.;
it is not an observer frame. To reposition the 
representation in an observer frame it must be
translated from an anti-symmetric representation
into a symmetric representation. This is analogous
to transforming homothetic curvature due to
torsion into 
rotation curvature. Once translated the 
stress-energy tensor for the free-field
must be added to the rotation curvature
coming from the source as both must now be regarded
as contributing to the rotation curvature. 

2. The second generalisation involves adding
a $U(1)$ phase factor related to the charged
source momentum with non-zero rest mass
to the symmetric part of the 
metric. This is `on the observer frame'
(i.e.$ds^2\neq0$) and
generates rotation curvature in a manner
analogous to the rotation curvature generated
in Gravitation theory. The potential 
generating a current in this
case is then the 4-momentum of the charged
electron or proton. This current is a 
conserved quantity;
\begin{equation}
-{i\over4}\Gamma^\alpha_{\gamma\alpha}
=\left(
-i\partial_{\gamma}e^{ip{\cdot}x}\right)
e^{-ip{\cdot}x}=p_{\gamma}
=j_{\gamma}
\end{equation}
so that clearly $\partial^{\gamma}j_\gamma=0$
when we treat momentum and position as independent
variables in the quantum representation.
The presence of a phase factor in the 
metric for an electron means that
\[
p^2=m_o^2e^{ip{\cdot}x}
\]
but if we reinterpret this in terms
of operators and wave-functions we have 
instead;
\[
\hat{p}^\mu\hat{p}^\nu
g_{\mu\nu}
=
\hat{p}^\mu\hat{p}^\nu
\eta_{\mu\nu}
<x\,|\,p>
=m_o^2<x\,|\,p>
\]
Thus at short distance scales (there is an
intrinsic factor of $\hbar$ in the phase
factor) space-time
takes on a wave-nature which means that a 
classical particle description is no
longer appropriate. 
In particular the non-zero zero point 
energy present in the stress-energy tensor
means that we are dealing with an
harmonic oscillator with non-zero
minimum energy. In the framework of
general relativity the wave part of 
the wave-particle duality
is thus due to space-time curvature
at short scales; it's another way of 
looking at the world. Unfortunately the
 precise quantitative difference
in the
scales of the coupling parameters for electro-magnetism
and gravitation has not been given an explanation but
it has been noted that, prior to translation, the
difference is infinite! It is thus perhaps no surprise
that there results with translation a large difference
(at least at low energies).

After translating the free-field stress-energy
tensor to a symmetric form we may consider
it a component part of the rotation curvature
generated by metric (\ref{sourcemetric}) instead of
homothetic curvature. 
With this proviso summing the contributions to rotation
curvature arising from the metric 
(\ref{sourcemetric}) we may write;
\begin{eqnarray}
&-&
{1\over{\kappa^2}}
(R^{\gamma\delta} 
-{1\over2}\eta^{\delta\gamma}R)
=
{\alpha\over2}.
m_o
U^{\gamma}U^{\delta}
{d\tau\over{dx^o}}\delta^3(x^i-x^i_p
\tiny{(\tau)})
\nonumber\\
&+&F^{\gamma\,\alpha}
F_{\alpha}^{\;\;\delta}+{1\over4}
\eta^{\gamma\,\delta}
F^{\mu\,\nu}
F_{\mu\,\nu}
=T^{\gamma\delta}_P+T^{\gamma\delta}_F
\label{final}
\end{eqnarray}
where  the
zero-point energy has been discarded
as is usually done in quantum theory.
The covariant derivative of both sides of
eq.(\ref{final}) must vanish which gives
us the inhomogenous Maxwell equations.
The homogenous Maxwell equations result 
from considering the free-field alone
(i.e. equation (\ref{set})) by setting
source 4-momentum to zero. The classical
equations of motion for a point particle
in an electro-magnetic field result
if we substitute the ordinary derivative for
the covariant derivative. Thus we can recover
classical electro-dynamics.
 Also note that
we can also add to the right side of 
eq.(\ref{final})
the contributions from gravitation by letting the
symmetric part of the metric vary with the
gravitational potentials. It will of course
enter with the opposite sign to the contributions
from the electromagnetic field. For macroscopic
charged matter we would of course have to integrate
up the source term in eq.(\ref{final}). An object
with opposite charge will, however, enter with
opposite sign if we set 
$\eta_{\alpha\beta}e^{ip{\cdot}x}$
= unit negative charge and
$\eta_{\alpha\beta}e^{-ip{\cdot}x}$
as unit positive charge say.

 In forming
a combined graviational and electromagnetic
curvature equation it must however be remembered 
that the curvature is no longer purely
gravitational in nature. At the micro-scale
of space-time there is severe curvature
due to mass carrying electric  charge which is not
gravitational in nature. 

Does Einstein's metric theory of Gravity remain
intact under the above derivation of the 
electro-magnetic stress-energy tensor? Does the
equivalence principle still hold? 

To answer these questions requires some 
interpretation of the above derivation and 
metric (\ref{metric}). Noting that the 
derivation of the free-field electro-magnetic
stress-energy tensor was carried out with
a completely antisymmetric metric and that;
\[
g^A_{\alpha\beta}\;dx^{\alpha}dx^{\beta}=ds^2=0
\]
we can interpret the antisymmetric part of
metric (\ref{metric}) as a `co-moving' metric
in the light-cone frame of the photon; the null geodesic.
As noted above, one consequence of this is that
Lorentz scalars for macroscopic matter
remain invariant under the total 
metric (i.e. the combination of
symmetric and anti-symmetric parts)
and we avoid the sort of problems 
related to measuring rods and clocks that 
led to so much criticism of the Weyl/Eddington 
theory. (I believe that at one stage Einstein
himself attempted to construct a version of
the Weyl/Eddington theory `on the light
cone'  to avoid the associated measurement
problems \cite{russian}).
The torsion I interpret as an essentially `local'
phenomena in the vicinity of each individual photon.
(See \cite{four}\cite{Hehl} for similar ideas).

For $g=g_a + g_s$ the vanishing of the covariant
derivative of the metric employed in the derivation
of the connection means that the (strong)
equivalence principle applies to the 
symmetric part of the metric which leads to
gravitation theory. The additional presence of
a phase factor at the quantum scale will be
expected to vanish for bulk matter as might
be expected in the classical limit. Thus
the classical theory of general relativity
remains intact.

Notice that (using eq (\ref{deriv}) and the definition of
$A^{\mu}$);
\[
\sigma^{\alpha\beta}_{\;\;\;\;,\alpha}
=-i\sigma^{\alpha\beta}_{\;\;\;\;,\gamma}
\,\sigma^{\gamma}_{\;\;\alpha}
=+i\sigma^{\gamma}_{\;\;\alpha}
\sigma^{\alpha\beta}_{\;\;\;,\gamma}
=-\sigma^{\alpha\beta}_{\;\;\;\;,\alpha}
\]
so that $\partial_{\mu}A^{\mu}=0$ and the
connection imposes the Lorentz condition.
This is the constraint in Proca's equation
which allows torsion for $m\neq0$ \cite{four} which the
co-moving metric makes implicit at $m=0$.


\begin{thebibliography}{9}
\bibitem{Weyl} H Weyl. Gravitation and
Electricity. 1918
(English translation;
'The Principle of Relativity ..."
Dover, New York. 1952. pp200-216)
\bibitem{A&P}
V.C. deAndrade and J.G Pereira
Torsion and the Electro-magnetic Field
gr-qc/9708051
see also
gr-qc/9706070 by same authors.
\bibitem{CM}
C. Moeller Mat.Fys.Skr.Dan.Vid.Selsk.
1,1-50(1961)
\bibitem{YM}
Y.M. Cho Phys. Rev. D 14, 2521-2525 (1976)
\bibitem{russian} Unified Field Theories.
V. Vizgrin. Birkhauser-Verlag. 1994
\bibitem{Klein}
O. Klein. Quantum theory and five dimensional
theory of relativity.
Z. Phys. 37 (1926) 895-906
\bibitem{UFT}A Einstein. The meaning of 
Relativity. 5th Ed. Princeton Univ. Press.
\bibitem{four} P von der Heyde et al
On Gravitation in Microphysics..
Proc. First Marcel Grossman meeting.
Nth Holland. 1977 255-278.
\bibitem{Hehl}
General Relativity with spin and torsion;
Foundations and prospects.
W. Hehl et al.
Reviews of Modern Physics.
Vol.48, No.3, pp393-416. 1976.
\bibitem{WE}
see for example;
K Borchsenius GRG 7, 527 (1976);
J Moffat Phys.Rev.D 15, 3520 (1977);
R McKellar Phys.Rev.D 20, 356 (1979);
A Jakubiec et al Lett.Math.Phys.9, 1 (1985);
M Ferraris et al GRG 14, 37 (1982);
K Horie hep-th/9506049.
\bibitem{T}
A Trautman; On the Structure of Einstein-Cartan
Equations. Symp. Mathematica. 12, 139 (1973)
\bibitem{Bilby}
K Kondo; Proc.2nd.Jap.Nat.Congress for Applied Mech,
41 (1952);
B Bilby et al; Proc.R.Soc.Lond; A231, 263 (1955)
\bibitem{one}
E Cartan;
Sur une generalisation de la
notion de courbure de Riemann et las espaces a torsion.
Acad.Sci., Paria
Comptes Rend.
174, 593-595
\bibitem{KK}
T. Kaluza.
`On the unity problem of physics'
English translation;
Modern Kaluza-Klein Theories.
T. Appelquist et al.
Addison-Wesley.  p61.(1981)
\bibitem{FP}
F.W. Hehl et al.
Gen. Rel. with spin and torsion; Foundations
and Prospects. Rev. Mod. Phys. 48. No.3 393-416 (1976).
\bibitem{F.W.}
F.W.Hehl et al. Physics Reports 258  1-171 (1995)
\bibitem{3}
D Sciama; "On the analogy between charge and
spin in general relativity" in Recent
Developments in General Relativity
Pergammon+PWN, Oxford. 415. (1962) and
Rev.Mod.Phys. 36, 463 and 1103 (1964)
\bibitem{4}
T Kibble; J.Math.Phys.2,212 (1961)
\bibitem{Mielke}
E Mielke et. al.;
Yang-Mills configurations from 3D R-C geometry.
Physics Let. A 192  153-162 (1994)
and contained references.
\bibitem{Lunev}
F Lunev;
Phys.Lett.B;
295, 99 (1992);
311, 273 (1993);
314, 21 (1994)
and Theor.Math.Phys.
94, 48 (1993)
\bibitem{U}
A Unzicker; e-archive gr-qc/9612061
\bibitem{Rad}
V Radovanovic et. al.;
Space-time geometry of 3d Yang-Mills theory.
Call. Quantum Grav. 12, 1791-1800 (1995)
\bibitem{Ham}
L. Garcia de Andrade et al.
Einstein-Cartan-Proca Geometry.
Gen. Relativity and Gravitation.
Vol 27, No. 12 1995 p1259
\bibitem{Kaku}Quantum Field Theory. M.Kaku.
Oxford 1993 p101.
\bibitem{two}
P.A.M. Dirac. 
General Theory of Relativity.
Princeton University Press. 1996.p.21.
\bibitem{Eddington}
A. Eddington. 
A Generalisation of Weyl's theory of
the Electromagnetic  and Gravitational field.
Proc. Royal Soc. Lond. A., 99. pp104-122. (1921)
\bibitem{CFT}
see for example F. Low. Classical Field Theory.
pp 271-277. Wiley and Sons. 1997
\bibitem{three}
See for example  Quantum Field Theory. M.Kaku.
Oxford 1993 pp28-30.
\bibitem{L}
S.L.Glashow et al. Phys.Rev.D 56, 2433. (1997),
R.B.Mann et al. Phys.Rev.Lett. 76, 865. (1996)
\bibitem{five}
C. W. Misner et. al.
Gravitation. Freeman and co. (1973).
\bibitem{GF}
G.R.Filewood.
Proc. Fifth Marcel Grossman Meeting.
World Scientific. 827 (1989)
\end{thebibliography}
\end{document}